\documentclass[
 reprint,
 superscriptaddress,
 amsmath,amssymb,
 aps,
 prx,
 floatfix
]{revtex4-2}

\usepackage{graphicx}
\usepackage{dcolumn}
\usepackage{bm}
\usepackage{hyperref}
\usepackage{booktabs}
\usepackage{multirow}
\usepackage{tabularx}
\usepackage{subcaption}
\usepackage{placeins}
\usepackage{float}
\usepackage{xcolor}
\usepackage{footnote}
\usepackage{booktabs, caption}
\usepackage{siunitx}
\usepackage[flushleft]{threeparttable}

\begin{document}

\title{A Transcorrelated Wave-Function Framework for Solids: An Application to Bulk and Defected Silicon}

\author{Kristoffer Simula}
\affiliation{Max Planck Institute for Solid State Research, Heisenbergstr. 1, 70569 Stuttgart, Germany}
\author{Johannes Hauskrecht}
\affiliation{Max Planck Institute for Solid State Research, Heisenbergstr. 1, 70569 Stuttgart, Germany}
\author{Evelin Martine Corvid Christlmaier}
\affiliation{Max Planck Institute for Solid State Research, Heisenbergstr. 1, 70569 Stuttgart, Germany}
\author{Pablo Lopez-Rios}
\affiliation{Max Planck Institute for Solid State Research, Heisenbergstr. 1, 70569 Stuttgart, Germany}
\author{Daniel Kats}
\affiliation{Max Planck Institute for Solid State Research, Heisenbergstr. 1, 70569 Stuttgart, Germany}
\author{Denis Usvyat}
\affiliation{Institut f\"ur Chemie, Humboldt-Universit\"at zu Berlin, Brook-Taylor-Str. 2, Berlin 12489, Germany}
\author{Ali Alavi}
\affiliation{Max Planck Institute for Solid State Research, Heisenbergstr. 1, 70569 Stuttgart, Germany}
\affiliation{Yusuf Hamied Department of Chemistry, University of Cambridge, Lensfield Road, Cambridge CB2 1EW, UK}
\date{\today}

\begin{abstract}
Accurate wave-function descriptions of pristine and defected solids remain challenging due to the simultaneous presence of finite-size, basis-set, and correlation errors. While embedding techniques alleviate finite-size effects and correlated wave-function approaches systematically improve correlation, basis-set incompleteness continues to limit practical accuracy. Here we present a study of transcorrelated (TC) many-body wave-function methods on properties of solid state systems. We augment the existing xTC theory to periodic systems, and establish an unified transcorrelated embedding framework that integrates periodic TC theory with fragment-based correlated solvers. Using silicon as a test case, we validate the method against coupled-cluster, FCIQMC, and diffusion Monte Carlo benchmarks for bulk. Then we apply TC embedding to calculation of formation energies of two silicon self-interstitials. The TC Hamiltonian yields rapid basis convergence and quantitatively reliable defect formation energies at the triple-$\zeta$ level, substantially reducing the basis-set bottleneck for wave-function treatments of crystalline defects.
\end{abstract}

\maketitle
\section{Introduction}

The accurate description of electronic structure in solids, especially in presence of point defects, remains one of the central challenges of theoretical condensed matter physics. The complexity arises from the simultaneous need to control three sources of error: (i) finite-size effects (FSE) in supercell models, (ii) basis-set incompleteness, and (iii) incomplete treatment of electron correlation. Wave-function methods offer systematic improvability in (iii), but their steep scaling with system size and basis dimension makes it difficult to reach convergence in (i) and (ii)~\cite{simula2023nv,verma2023,ertekin2013,chen2023nv,chen2025hbn,gruber2018,masios2023}, particularly for point defects that require large simulation cells and often high basis set resolution. 

Because of these difficulties, wave-function based methods that directly solve the many-electron Schrödinger equation have seen limited use in solid-state physics, despite their systematic improvability and their success in molecular systems.  Density functional theory (DFT) remains the workhorse of condensed-matter simulation because of its favorable scaling and ability to treat large periodic systems.  However, DFT often lacks the accuracy required for quantitative predictions, particularly for systems where correlation effects are strong, motivating the search for alternatives that combine high accuracy with computational feasibility.

Diffusion Monte Carlo (DMC)~\cite{foulkes2001} solves the exact Schrödinger equation within the fixed-node approximation and has been successfully applied to solid-state systems~\cite{simula2023nv,ertekin2013}. Its favorable $N^{3}$–$N^{4}$ scaling with system size $N$ and small basis-set errors allow the treatment of supercells containing hundreds of atoms, establishing DMC as a standard benchmark for solids. However, the fixed-node approximation introduces an uncontrolled, trial-wave-function–dependent bias, and converging this bias with respect to simulation cell sizes for defects can be prohibitively expensive due to the need for, e.g., backflow or multideterminantal wave functions.

Second-quantized methods such as coupled-cluster (CC) theory and full configuration-interaction quantum Monte Carlo (FCIQMC)~\cite{booth2009,cleland2010,guther2020,ghanem2019,ghanem2020} provide a systematically improvable hierarchy toward the exact many-electron solution. However, their scaling with system size and basis set is far worse than that of DMC, making their use for realistic defect description with high resolution infeasible. 

Embedding techniques have therefore emerged as a powerful strategy to alleviate finite-size effects~\cite{masur2016,lin2020,fciqmc-fragment,schaefer21,Berkelbach21,muechler2022,aperiodic} and enable the use of second-quantized methods for realistic systems with defects. In these approaches, a chemically motivated \emph{fragment} containing the defect is treated with a high-level correlated solver, while the surrounding crystal is described at the mean-field level with sufficiently large supercells or even with an entirely non-defective matrix using the aperiodic defect method \cite{aperiodic}. Embedding based on localized Wannier and projected atomic orbitals (PAOs) enables systematically enlargable fragments that capture essential defect physics at greatly reduced cost. 

Even within such frameworks, basis-set incompleteness remains a severe obstacle. Plane-wave bases offer periodicity and systematic convergence but require prohibitively large cutoffs for correlated solvers, whereas localized Gaussian-type orbitals, though compact, introduce incompleteness errors that are difficult to eliminate in periodic environments. As the Gaussian basis sets are enlargened, linear dependencies arise, often making convergence with respect to basis set impossible~\cite{vandevondele2007,peintinger2013}. Recent attempts have been made to cure the linear dependency issue and introduce heavier quadruple-zeta gaussian basis sets for solids that match plane-wave accuracy~\cite{lee2021}. Within a periodic MP2 framework, the basis set incompleteness issues have also been addressed using the explicitly correlated theories~\cite{usvyat2013,booth2013pw}. 

Yet, among the three principal sources of error the basis-set incompleteness may remain the main practical limitation for descriptions of defects under second-quantized theories. While embedding mitigates the finite-size effects and correlated solvers control the electron correlation, achieving a reliable and transferable basis convergence continues to limit attainable accuracy. These persistent limitations motivate the development of new theoretical frameworks with systematic treatment of correlation and small basis-set errors.

The transcorrelated (TC) framework~\cite{sakuma2006,ochi2012,haupt2025,simula2025ecp,simula2025tc_transition_metals,ammar2023,liao2021,schraivogel2021,lee2023,liao2023dmrg} represents a conceptually distinct route to accurate theoretical description of electronic-structure, directly addressing the basis-convergence problem by performing a similarity transformation of the Hamiltonian with respect to a Jastrow factor $J$ that captures the dominant cusp and dynamical correlation physics~\cite{drummond2004,cohen2019,haupt2023}. This transformation yields an effective Hamiltonian with compactified wave functions and greatly accelerated basis convergence, at the cost of introducing three-body interaction terms and non-Hermiticity. The recently developed xTC approximation replaces these three-body contributions with an effective two-body operator, enabling efficient and accurate many-body calculations~\cite{christlmaier2023}. When combined with norm-conserving pseudopotentials (PPs)~\cite{simula2025ecp}, the Jastrow can focus on valence correlations while avoiding nuclear cusps, further improving efficiency for large atoms~\cite{simula2025tc_transition_metals}, molecules, or solids.

In this work, we introduce a xTC-PP framework for periodic solids and formulate a unified \emph{transcorrelated embedding} framework. We demonstrate the potential of these ideas with case studies in pristine and defected silicon. Starting from periodic Hartree–Fock (HF) orbitals and norm-conserving pseudopotentials, we optimize Drummond–Towler–Needs–type Jastrow factors in variational Monte Carlo (VMC)~\cite{drummond2004,haupt2023} and construct the periodic xTC–PP Hamiltonian. In defect simulations we downfold the xTC-PP Hamiltonian to fragment subspaces spanned by localized occupied Wannier functions and projected virtual orbitals. The resulting fragment Hamiltonians retain Jastrow-induced correlations between fragment and environment while remaining tractable for post-HF solvers.

The bulk silicon calculations are done in an eight-atom cell, comparing CCSD, DCSD, CCSD(T), and CCSDT results with FCIQMC and DMC benchmarks across multiple Jastrow cutoffs and basis levels. We show that the xTC–PP formalism achieves near-complete basis-set convergence already at the triple-$\zeta$ level without linear dependency issues. This shows that TC substantially improves the Gaussian-basis frameworks for solids. Our results are comparable to DMC benchmarks and we find coupled cluster with triple excitations to reach very close to FCI result. 

After bulk Si validation, we apply the xTC-PP embedding scheme on the traditional problem of  defect formation energy estimation, a task requiring often large cells, accurate basis resolution and correlation description. We use as test systems the hexagonal (H) and split (X) silicon self-interstitials in $65$-atom supercells, analyzing the convergence of defect formation energies with fragment size and basis level. We find convergent fragment sizes and obtain estimations of formation energies that fit in the experimental range measured for the H-interstitial. The final formation energies are hence obtained with relatively  large simulation cells, and they are likely to be very close to convergence in both basis set resolution and correlated treatment. 

The remainder of the paper outlines the periodic xTC–PP formalism, details the embedding construction, and presents bulk and defect benchmarks demonstrating the accuracy and efficiency of the TC theory for pristine and defected solids.

\section{Theory}\label{sec:Theory}

Here we provide an overview of the periodic xTC-PP method and introduce xTC-PP embedding for applications to lattice defects. For a detailed description of TC theory, we refer the reader to Refs.~\cite{cohen2019,christlmaier2023,haupt2023,simula2025ecp}. For the HF embedding, we follow the approach described in Ref.~\cite{fciqmc-fragment}.

Throughout this section, orbital indices \(p,q,r,s,t,u\) denote the full one-particle basis.  Occupied orbitals are labeled \(i,j,k,l\). In the embedding decomposition, we partition the full orbital space into \emph{fragment orbitals} \(p_f,q_f,r_f,s_f\) and \emph{environment orbitals} \(p_e,q_e,r_e,s_e\), using the same notation for occupied orbitals within each subspace. For the integrals over orbitals \(p,q\) or \(p,q,r,s\) and an operator \(\hat{O}\) we write \(\hat{O}_{pq}=\langle p|\hat{O}|q \rangle\) and \(\hat{O}_{pqrs}=\langle pq|\hat{O}|rs \rangle\), with the bras and kets being the one-electron basis orbitals, $|p\rangle=\phi_p(\mathbf{r})$.

The similarity transformation of the Hamiltonian $\hat H$ (under PP approximation) with a Jastrow factor $J_\alpha$ (of particle positions, $J_\alpha(\mathbf{r}_1,\ldots,\mathbf{r}_N)$) with at most two-electron terms leads to the following second-quantized TC-PP Hamiltonian $\hat H_{\mathrm{TC-PP}}$:

\begin{align}
    \begin{aligned}
        \label{eq:simtransf}
\hat H^{\mathrm{xTC}-PP}
&= e^{-J_\alpha}\hat H e^{J_\alpha}
\xrightarrow[\mathrm{xTC\!-\!PP}]{\text{\cite{cohen2019,christlmaier2023,simula2025ecp}}}\\
&E^{xTC}_0+\sum_{pq} {h}^{\text{xTC}}_{pq}\, a_p^\dagger a_q
+\tfrac12 \sum_{pqrs} {W}^{\text{xTC-PP}}_{pqrs}\, a_p^\dagger a_q^\dagger a_s a_r 
\end{aligned}
\end{align}
with ${W}^{\text{xTC-PP}}$ being the two-body xTC-PP interaction term defined as
\begin{align}
    \begin{aligned}
    \label{eq:T_pqrs}
{W}^{\text{xTC-PP}}_{pqrs}
&= {V}_{pqrs} + \Delta V_{pqrs}, \\
\Delta V_{pqrs}&=  - {K}_{pqrs} + {P}_{pqrs}
+\Delta {W}_{pqrs}, \\
\Delta {W}_{pqrs}&= - \sum_{ij}\left({L}_{priqsj}
                                        -{L}_{priqjs}
                                        - {L}_{prijsq}\right)\gamma_{ij}
    \end{aligned}
\end{align}
and the one-body xTC term as
\begin{align}
    \label{eq:F_pq}
    \begin{aligned}
{h}^{\text{xTC}}_{pq} &= {h}_{pq} + \Delta {h}_{pq},\\
\Delta {h}_{pq}&= -\frac{1}{2}\sum_{ij}\left(\Delta {W}_{piqj}-\Delta {W}_{pijq}\right)\gamma_{ij},
\end{aligned}
\end{align}
and the operators $a_p^\dagger$ and $a_p$ are the creation and annihilation operators of the orbitals $\phi_p(\mathbf{r})$. 
The one-body reduced density matrix, $\gamma_{ij}$, is that of Hartree-Fock (HF) reference. We have defined \(\Delta{W}_{pqrs}\) and \(\Delta {h}_{pq}\) to be the  corrections to transcorrelated $1$- and $2$-body TC terms due to the xTC approximation.
Because of the xTC, we also introduce a correction to the constant energy term~\cite{christlmaier2023}: 
\begin{align}
    \label{eq:e0-xtc}
    \begin{aligned}
    E^{\text{xTC}}_0 &= E_0 + \Delta E_0^{\text{xTC}},\\
    \Delta E_0^{\text{xTC}}&=-\frac{1}{3}\sum_{ij}\Delta h_{ij}\gamma_{ij}.
\end{aligned}
\end{align}

In Eqs.~(\ref{eq:simtransf})-(\ref{eq:F_pq}), the one-electron operator $\hat h$ contains the kinetic energy and electron-nucleus pseudopotential parts. $\hat{V}$ is the bare two-electron Coulomb interaction $\hat{V}(\mathbf{r}_1,\mathbf{r_2})=1/|\mathbf{r}_2-\mathbf{r}_1|$. The $\hat K$, $\hat L$ and $\hat P$ operators originate from the Baker-Campbell-Hausdorff expansion of the similarity transformed Hamiltonian, giving rise to the kinetic energy commutators $\hat K+\hat L=\left[\frac{1}{2}\sum_p\nabla_p^2,J_\alpha\right]+\left[\left[\frac{1}{2}\sum_p\nabla_p^2,J_\alpha\right],J_\alpha\right]$, and the pseudopotential commutators $\hat P=\left[\sum_p\hat{V}_{ecp}^p,J_\alpha\right]+\left[\left[\sum_p\hat{V}_{ecp}^p,J_\alpha\right],J_\alpha\right]$. 
$\hat K$ is a two-body operator, while $\hat L$ and $\hat P$ are three-body operators. Based on earlier findings \cite{simula2025ecp,simula2025tc_transition_metals}, we neglect the three-body contribution of $\hat P$, which has proved to be a very good approximation.

We use a Jastrow factor $J_{\alpha_u,\alpha_\chi,\alpha_f}=\sum_{i\neq i}u_{\alpha_u}(\mathbf{r}_i,\mathbf{r}_j) + \sum_\text{I}\sum_\text{i}\chi_{\alpha_\chi}(\mathbf{r}_i,\mathbf{R}_\text{I}) + \sum_\text{I}\sum_{\text{i}\neq\text{j}}f_{\alpha_f}(\mathbf{r}_i,\mathbf{R}_\text{I},\mathbf{r}_j)$, defined by parameters $\alpha_u,\alpha_\chi,\alpha_f$, of Drummond-Towler-Needs type \cite{drummond2004} with two-body ($u$), one-body ($\chi$), and three-body ($f$) terms. In the Jastrow terms, $\mathbf{r}_i$ and $\mathbf{r}_j$ are the positions of the electrons, and $\mathbf{R}_\text{I}$ is the position of the nuclei. The parameters $\alpha_u,\alpha_\chi,\alpha_f$ are optimized with VMC. Each Jastrow term is truncated at a cutoff length, denoted by $(L_u,L_\chi,L_f)$. The $u$ term captures the electron-electron cusp condition. With periodic boundary conditions (PBC), we employ the periodic version of the Jastrow factor \cite{drummond2004}. The orbitals are obtained from a periodic HF calculation with a bare Hamiltonian.

We call the HF energy of the non-transcorrelated periodic system as the reference energy, or the non-TC reference energy. With TC, we call the expectation value of the TC Hamiltonian with respect to the HF wave function the transcorrelated, or xTC-PP reference energy.

When we study fully periodic bulk silicon system, without embedding, we use both non-TC $\hat{H}$ and transcorrelated $\hat{H}_{\text{TC-PP}}$ evaluated in the full Hilbert space of the chosen basis set to do coupled cluster (CC) theory and full configuration interaction quantum Monte Carlo (FCIQMC) to get the correlation energy of the supercell. We also evaluate the total energy of the system using diffusion Monte Carlo (DMC), with both Slater-Jastrow (SJ) and Slater-Jastrow-backflow (SJB) trial wave functions. 

A periodic mean-field embedding implies separation of the full system into a fragment and an environment. The fragment is treated with a correlated wave function method, while the environment is left at the mean-field level. The embedding technique is described in detail in Ref.~\cite{fciqmc-fragment}. The fragment Hamiltonian is 
\begin{align}
    \begin{aligned}
        \label{eq:Hfrag_total}
    \hat{H}_\text{frag} 
    =& \sum_{p_f q_f} h^\text{frag}_{p_f q_f} a_{p_f}^\dagger a_{q_f} \\
    &+ \tfrac12 \sum_{p_f q_f r_f s_f} {V}_{p_fq_fr_f s_f} a_{p_f}^\dagger a_{q_f}^\dagger a_{s_f} a_{r_f} 
    + E^\text{frac}_0 ,
    \end{aligned}
\end{align}
with 
\begin{align}
    \begin{aligned}
        \label{eq:hfrag}
    h^\text{frag}_{p_f q_f}
    =& h^\text{per}_{p_f q_f} + \sum_{i_e} \left[2{V}_{p_f i_e q_fi_e} - {V}_{p_f i_e i_e q_f}\right]\\
    =& f^\text{per}_{p_f q_f} - \sum_{i_f} \left[2{V}_{p_f i_f q_fi_f} - {V}_{p_f i_f i_f q_f}\right],
    \end{aligned}
\end{align}
% In the above two equations, we have used the notation
% \begin{equation}
% (pq|rs) = \iint d\mathbf{r}_1 d\mathbf{r}_2 \phi_p^*(\mathbf{r}_1) \phi_q(\mathbf{r}_1) \frac{1}{|\mathbf{r}_1-\mathbf{r}_2|} \phi_r^*(\mathbf{r}_2) \phi_s(\mathbf{r}_2) .
% \end{equation}
where the one-electron Hamilotonian $\hat h^\text{per}$ and the Fock operator $\hat f^\text{per}$ correspond to the complete periodic system. Finally, in order to reproduce the periodic HF energy per cell $E^\text{per}_\text{HF}$ in $E^\text{frag}_\text{HF}$ we define $E^\text{frac}_0$ as
\begin{equation}
    \label{eq:Efrac_nuc}
    E^\text{frag}_0 = E^\text{per}_\text{HF} - 2\sum_{i_f} h^\text{frag}_{i_f i_f} - \sum_{i_f j_f} \left(2{V}_{i_f j_fi_f j_f} - {V}_{i_f j_fj_f i_f}\right).
\end{equation}

Before defining the fragment, the HF orbitals of the full periodic system are localized. The localization of the occupied Wannier orbitals is carried out using the method of Refs. \cite{zicovichwilson2001, ZD_WHF07}. For the virtual manifold we employ projected atomic orbitals (PAOs) \cite{Usvyat2010}. The fragment is then defined by a set of "seed" atoms, defining which Wannier orbitals belong to the fragment on the basis of their Mulliken populations~\cite{fciqmc-fragment}. As PAOs are non-orthogonal and even redundant, the PAOs belonging to the fragment are canonicalized with the cutoff threshold for the eigenvalues the PAO-overlap matrix of 10$^{-4}$. 

To incorporate the transcorrelated embedding in the fragment's one-electron Hamiltonian 
we append it with the environment's mean-field contributions from the Jastrow commutator
operators:
\begin{align}
    \begin{aligned}
        \label{eq:h-xTC-frag}
    h^\text{xTC-PP-frag}_{p_f q_f}
    =& h^\text{frag}_{p_f q_f} + \Delta h_{p_f q_f}\\
       & +\sum_{i_e} \left[2\Delta V_{p_f i_e q_fi_e} - \Delta V_{p_f i_e i_e q_f}\right].
    \end{aligned}
\end{align}
The two-electron part of the fragment Hamiltonian takes the form 
\begin{align}
    \begin{aligned}
        \label{eq:2-body-xTC-frag}
    {W}^{\text{xTC-PP-frag}}_{p_fq_fr_fs_f} =& V_{p_fq_fr_f s_f} + \Delta V_{p_fq_fr_fs_f}
    \end{aligned}
\end{align} 
%the one-electron term as 
%\begin{align}
%    \begin{aligned}
%        \label{eq:h-xTC-frag}
%    h^\text{xTC-PP-frag}_{p_f q_f}
%    =& h^\text{frag}_{p_f q_f} + \Delta h_{p_f q_f}\\
%       & +\sum_{i_e} \left[2\Delta V_{p_f i_e q_fi_e} - \Delta V_{p_f i_e i_e q_f}\right],
%    \end{aligned}
%\end{align}
Finally the constant energy term is redefined as
\begin{align}
    \begin{aligned}
        \label{eq:Efrac_nuc-xtc}
        E^\text{xTC-PP-frag}_0 =& E^\text{frag}_0 + 2\sum_{i_e} \Delta h_{i_e i_e} \\
        &+ \sum_{i_ej_e}\left[
            2\Delta V_{i_e j_e i_ej_e} - \Delta V_{i_e j_e j_ei_e}
         \right] \\
         &+ \Delta E_0^{\text{xTC}}
    \end{aligned}
\end{align}
such that the expectation value of the xTC-Hamiltonian $\hat{h}^\text{xTC-PP-frag}+\hat{W}^{\text{xTC-PP-frag}}+E^\text{xTC-PP-frag}_0$ within the fragment's occupied space reproduces the periodic xTC-PP reference energy, defined as the expectation value of the transcorrelated Hamiltonian of the fully periodic system with respect to the periodic HF wave function. This energy, which we denote as $E_{\text{HF}}^\text{xTC-PP-per}$,  serves as the reference energy for the transcorrelated fragment treatment within the xTC-PP embedding model.
% \begin{align}
%     \begin{aligned}
%         \label{eq:Efrac_nuc-xtc}
%         E_{\text{HF}}^\text{xTC-PP-frag} = E_{\text{HF}}^\text{xTC-PP-per}.
%     \end{aligned}
% \end{align}

We also note that in the formulated above xTC-embedding model the transcorrelated-enviroment is coupled to the fragment not only via the mean field of the $\Delta \hat{V}$ operator included in $\hat{h}^\text{xTC-PP-frag}$, but also from the $\hat{L}$ operator contributions to $\Delta {W}_{p_fq_fr_fs_f}$ and $\Delta h_{p_f q_f}$, as the sums over $ij$ in eqs. (\ref{eq:T_pqrs}) and (\ref{eq:F_pq}) run over {\it all} occupied orbitals, including those in the environment. 
%This introduces the explicit correlation between fragment and the environment via the Jastrow interaction. 
The integrals of the $\hat{K}$, $\hat{P}$ and $\hat{L}$ operators are calculated numerically under the minimum image convention using the $\Gamma$-point Bloch sums of the Wannier functions $i_e$ and fragment orbitals $p_f$, $q_f$. Yet, the Coulomb integrals $V_{p_fq_fr_fs_f}$ used in eqs. (\ref{eq:hfrag}), (\ref{eq:Efrac_nuc}) and (\ref{eq:2-body-xTC-frag})  are evaluated in the direct space using the periodic local density fitting as described in Ref. \cite{fciqmc-fragment}.  

With the obtained second-quantized transcorrelated fragment Hamiltonians, we use CC theory to obtain fragment correlation energies. The total energy of the full system is then obtained as the HF energy of the periodic cell plus the CC correlation energy of the fragment. 

We estimate the defect formation energies as
\begin{equation}
   E_f = E_\text{HF}^\text{defect} - \frac{N_a^d}{N_a^b}E_\text{HF}^\text{bulk} 
            + E^\text{defect}_\text{corr} - \frac{N_e^d}{N_e^b}E_\text{corr}^\text{bulk},  
\end{equation}
where $E_\text{HF}^\text{defect}$ and $E_\text{HF}^\text{bulk}$ are the fully periodic HF total energies of the defect and bulk supercells, respectively; $E_\text{corr}^\text{defect}$ and $E_\text{corr}^\text{bulk}$ are the corresponding fragment correlation energies; $N_a^d$ ($N_a^b$) is the number of atoms in the defect (bulk) supercell; and $N_e^d$ ($N_e^b$) is the number of electrons in the defect (bulk) fragment. This definition of the formation energy asymptotically approaches the full simulation cell formation energy with increasing fragment size.

\begin{figure}[H]
    %\centering
    \includegraphics[scale=.15]{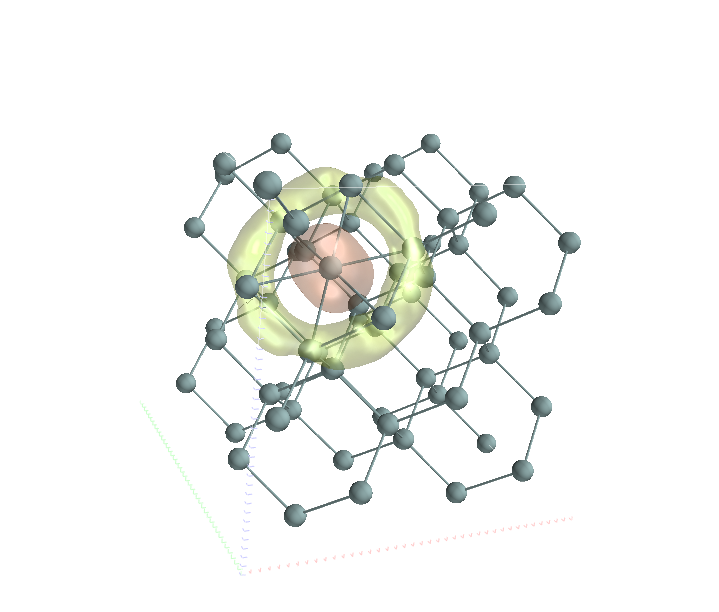}
    \includegraphics[scale=.15]{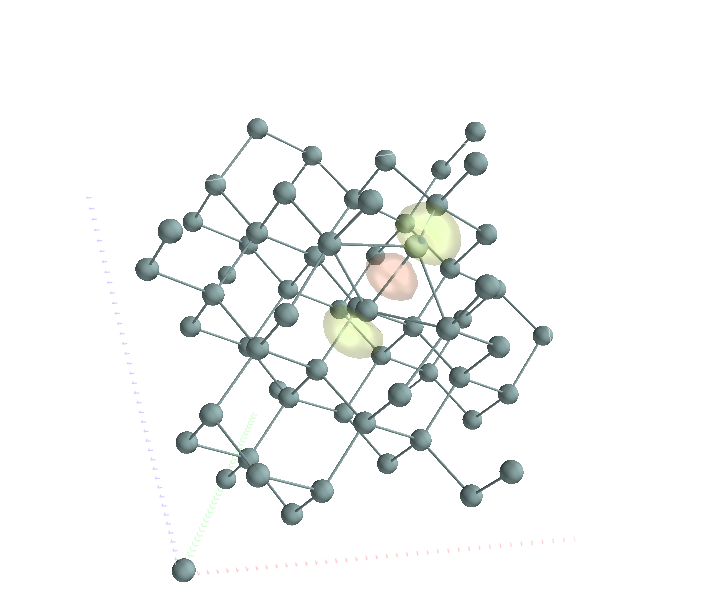}
    
    \caption{\label{fig: defect illustration}The periodic simulation cells of the relaxed hexagonal (H,left) and split (X, right) silicon self-interstitial defects used in the formation energy calculations,  with the highest  occupied fragment orbitals plotted.}
\end{figure}

\section{Computational Details}

\begin{figure*}[t]
    \includegraphics[scale=1.3]{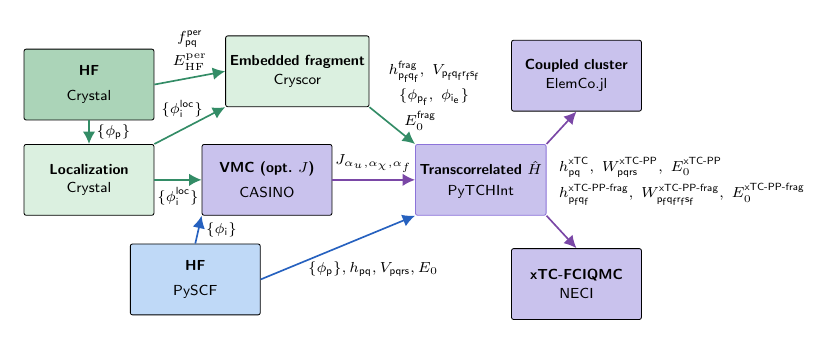}
\caption{\label{fig:workflow}
Computational workflows used in this work. Boxes represent the calculation stages and the codes employed. Directed arrows indicate the flow of data between stages, and their labels specify the quantities transferred (see Sec.~\ref{sec:Theory}). The orbitals $\phi_p$ denote periodic HF orbitals, $\phi_i^\mathrm{loc}$ their localized counterparts, and $\phi_{p_f}$ and $\phi_{i_e}$ the $\Gamma$-point Bloch sums of the Wannier functions $i_e$ and fragment orbitals $p_f$, $q_f$, respectively. $E_0$ is the nuclear repulsion energy. Green indicates steps specific to the embedding workflow, blue those specific to fully periodic calculations, and lavender the stages shared by both workflows. For clarity, we show the xTC data passed to CC and FCIQMC only once.}
\end{figure*}

We separate the study of transcorrelated solid-state theory into two cases: bulk silicon in the periodic $8$-atom conventional cell, and the silicon self-interstitial defects in periodic $65$-atom supercells. For the bulk system, we model the full periodic conventional cell at all stages of our workflow. For the defect, we perform a periodic HF calculation, localize the resulting orbitals, and define a periodic fragment on which the correlation treatment is focused. Figure~\ref{fig:workflow} summarizes the workflow for these two cases.

Both workflows begin with a periodic HF calculation. In the bulk case, the occupied HF orbitals are used to construct the Slater–Jastrow trial wave function for the VMC optimization. The subsequent evaluation of xTC–PP integrals employs the optimized Jastrow factor together with the HF orbitals and the Hamiltonian matrix elements obtained at the HF stage.

In the embedding workflow, the HF orbitals are first localized and then used to build the trial wave function for VMC. The xTC–PP integral calculation receives the fragment orbitals, the occupied environment orbitals, and the corresponding Hamiltonian elements produced during the fragment-construction phase.

In the following, we describe each step of the workflows in detail.

In bulk, we study the convergence of total energy with respect to basis size and level of correlation treatment. In defected systems we calculate formation energies and compare against theoretical and experimental benchmark values. The defects studied are hexagonal (H) and split (X) defects. The defect simulation cells are constructed by adding one interstitial Si atom to the bulk $64$-atom supercell. The lattice constant is taken to be the experimental value of $5.43$\,\AA in all cases.

The H defect is constructed by inserting a single atom in the center of a hexagonal ring of silicon atoms. The split defect consists of two symmetrically equivalent atomic positions on both sides of an atomic site of a pristine lattice. The atomic positions of the periodic defect supercells are relaxed using DFT with HSE06 hybrid functional~\cite{hse06}. We used plane-wave basis and PAW treatment with the VASP package \cite{kresse1996} for relaxations. The VASP relaxations use a plane-wave cutoff of $400$\,eV and a $3\times3\times3$ Monkhorst-Pack k-point grid. The relaxed defect structures are then used in all subsequent calculations. The relaxed defect structures are shown in Fig.~\ref{fig: defect illustration}.

 We use the periodic HF module of PySCF~\cite{sun2020} with a single $\mathbf{k}$-vector at $\Gamma$ to get the bulk $8$-atom cell orbitals. The HF orbitals of the 64/65-atom supercells (64 for bulk and 65 for defects) are obtained with the Crystal17 package~\cite{dovesi2018}, using a $3\times 3\times 3$ Monkhorst-Pack k-point grid. The orbitals are generated using ccECP pseudopotentials and corresponding cc-pVDZ (DZ) and cc-pVTZ (TZ) basis sets for silicon~\cite{ccecp2022}. Both PySCF and Cystal17 HF calculations use the same ccECPs and basis sets. To avoid linear dependencies we remove long-tailed Gaussians with exponents smaller than $0.08$ a.u. 

With the HF orbitals, we construct a trial wave function with a single Slater-determinant to optimize the Jastrow parameters with VMC using the CASINO package \cite{needs2020}. The optimization is done by minimising the variance of the transcorrelated reference energy \cite{haupt2023}. We use ($8,8,24$) parameters for the $u$, $\chi$, and $f$ terms, respectively. For the $8$-atom bulk cell we test two sets of Jastrow cutoffs, $(L_u,L_\chi,L_f)=(4,1,1)$ and $(5,3,3)$\,bohr, denoted as 4\,1\,1 and 5\,3\,3 in the following sections. In the TZ basis, we show that these two cutoffs lead to very similar total energies for the 8-atom cells. In the scope of the present study, for the larger defect cells, we use only the 4\,1\,1 Jastrow. 

The optimized Jastrow parameters are used together with the orbital and Hamiltonian information to construct the xTC-PP integrals numerically according to Eqs.~\ref{eq:T_pqrs},~\ref{eq:F_pq},\ref{eq:h-xTC-frag}, and \ref{eq:2-body-xTC-frag} with our in-house code, PyTCHInt, adapted to periodic calculations. For formation energies, we use the Jastrow optimized in the defect supercell for the bulk fragments for balanced energy comparison. 

When computing numerical xTC-PP integrals of Si conventional cell, we use all of the orbitals obtained in the HF phase with a given gaussian basis. In fragment calculations, we include all fragment orbitals and the occupied environment orbitals in the integral evaluation. This folds the electron correlation at the xTC level between the fragment and the environment into the fragment Hamiltonian. After evaluating the xTC-PP integrals for the fragment and occupied environment orbitals, we freeze the environment electrons and hence fold the xTC-PP $2$- and $1$-body contributions from the environment into the fragment Hamiltonian. The periodic HF orbitals from PySCF and the  $\Gamma$-point orbital Bloch sums from Cryscor are passed via a molden-format interface (supplemented with simulation cell lattice vectors) to PyTCHInt.

We use the FCIDUMP-format \cite{Knowles89} to pass the ${h}_{pq}$ and ${V}_{pqrs}$ of the $8$-atom conventional cell and the direct space fragment integrals $h_{p_fq_f}$ and $V_{p_fq_fr_fs_f}$ evaluated for the defect and corresponding bulk supercells.  The fragment Hamiltonian construction is done as implemented in the Cryscor package \cite{fciqmc-fragment}.

Instead of using atom-centered grids for integral evaluation as in molecular TC \cite{haupt2023,christlmaier2023,simula2025ecp,simula2025tc_transition_metals}, we use a uniform grid in the periodic cell with a density that converges the energies. We found the uniform grids to converge with fewer grid points than atom-centered grids for the $8$-atom silicon cell, based on a series of tests with xTC-PP-CCSD done in the $8$-atom bulk cell. Figure~\ref{figure: tc_integration_grid} and table \ref{tab:xtc_mp2_grid_tz_411} in the Appendix show that with atom-centered Becke grids $227\ 000$ grid points is needed to reach within $0.1$mE$_h$ of the energy per primitive cell obtained with $64\ 000$ uniform grid points. The computational cost of the xTC-PP integrals scales as $N_{\mathrm{grid}}^2$, and hence the use of uniform grids can lead to significant computational savings. We believe it is the use of PPs which allows the use of relatively sparse uniform grids for the evaluation of the xTC-PP matrix elements, as the strong cusps at the nuclei are removed, and the need to evaluate tight core orbitals as well as highly oscillatory valence orbitals (both of which require dense grids) is avoided.

\begin{figure*}[t]
    \centering
    \includegraphics[scale=.0915]{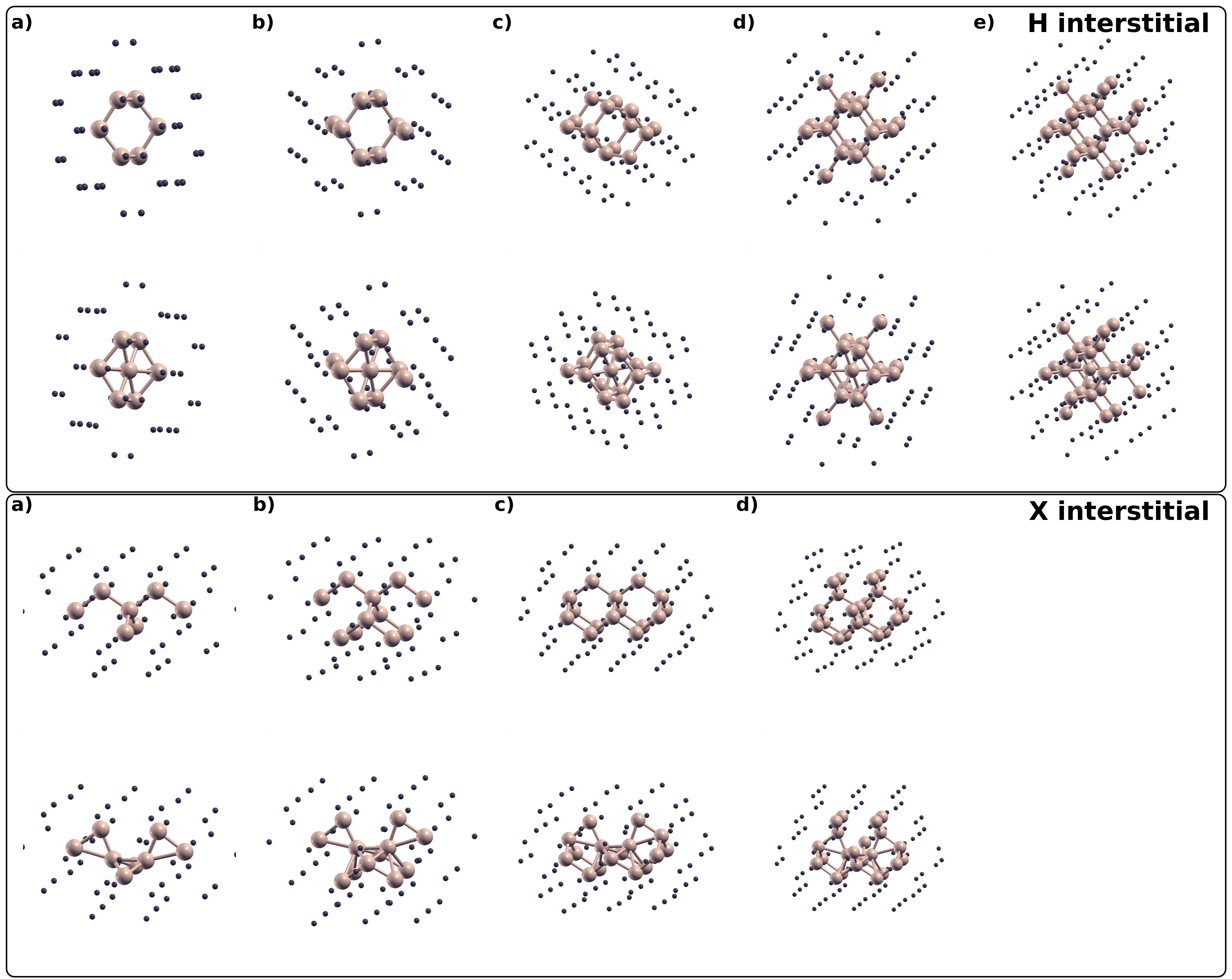}
    \caption{\label{figure: fragments}Embedding fragments used in this work. Each letter denotes a bulk–defect fragment pair: for the H-interstitial series (a–e) correspond to pairs with (7), (9), (15), (21), and (27) atoms in the defect fragments; for the X-interstitial series (a–d) correspond to pairs with (8), (12), (16), and (22) atoms. In bulk fragments corresponding to a certain defect fragment there is always one atom less. Fragment Si atoms are shown in orange; environment Si atoms are shown as smaller midnight-blue balls. For clarity, we only show a subset of the atoms in the periodic simulation cells of Fig. \ref{fig: defect illustration}.}
\end{figure*}

To study size-dependency of the embedding we define a series of fragments increasing in size around the defect interstitial atoms. We take the first (and second) nearest-neighbour atoms around the H (X) defect into the smallest fragment, and then keep adding shells of next-nearest-neighbour atoms to form larger fragments. The fragments used in this work are illustrated in Fig.~\ref{figure: fragments}, where, for clarity, we only depict the fragment atoms and part of the surrounding environment atoms in the periodic simulation cell. Fragment-atom counts range from $7$ to $27$ atoms for the H-interstitial series and from $8$ to $22$ atoms for the X-interstitial series. For each defect fragment we define a corresponding fragment in pristine bulk $64$-atom periodic supercell, with one atom less than in the defect fragment. 

Finally, with both the bare and xTC–PP second-quantized Hamiltonians we do correlation calculations.  In conventional $8$-atom bulk periodic silicon simulation cell we use the full Hamiltonian in xTC-PP-CC and -FCIQMC calculations, and in periodic defect supercells we use the embedding Hamiltonian to do xTC-PP-CC. We are not doing xTC-PP-FCIQMC for the defect fragments, as xTC-PP-CC is believed to provide sufficient accuracy, but this would be straightforward with the current implementations. The FCIQMC calculations are done using the NECI code~\cite{neci}, and coupled cluster calculations using the ElemCo.jl package \cite{elemcojl}. We do CC with singles and doubles (CCSD), CC with perturbative triples (CCSD(T)), full triples (CCSDT) and distinguishable cluster with singles and doubles (DCSD) calculations~\cite{kats2013,kats2014,kats2024}. 

For the non-TC CCSD(T) we present the complete-basis set (CBS) estimate of the non-TC CCSD(T) energy, based on a two-point extrapolation from the DZ and TZ results, in Fig.~\ref{fig: total energies} and Table~\ref{tab:etot-bulk}. We have assumed an exponential scaling in the HF energy in the basis set error as $E_\text{n}=E_{\infty} + B\text{exp}(-Cn)$, with $C=1.65$, as suggested by Jensen~\cite{jensen2005}. For the correlation energy we used the standard extrapolation, assuming scaling to follow a power-law dependent on the basis level.

NECI calculations are done with the adaptive-shift initiator FCIQMC (AS-FCIQMC)~\cite{ghanem2019,ghanem2020} using an initiator threshold of $10$, which yields faster convergence than using threshold of $5$, but we tested both to provide same total energies to within 1mE$_h$. 
We increased the walker populations in FCIQMC runs until convergence in total energies, with final populations ranging from $2\times 10^7$ to $4\times 10^8$ walkers depending on the basis and TC case.

In DMC calculations we use cutoffs of $(L_u,L_\chi,L_f)=(5,3,3)$\,bohr for the electron-electron, electron-nucleus, and electron-electron-nucleus Jastrow and backflow terms. We do two separate DMC runs with time steps of $0.02$ and $0.005$\,a.u., with corresponding walker populations of $2000$ and $8000$ walkers, respectively, and extrapolate linearly to zero time step. The trial wave function for DMC is the HF determinant obtained in TZ gaussian basis, using the same orbitals as the second-quantized methods used for the conventional $8$-atom conventional simulation cell.

To analyse the basis set incompleteness errors in the silicon interstitial problem between different core treatments, basis sets and Hamiltonians in the HF level, we in addition describe in Appendix a number of tests on the total and formation energies of silicon interstitials in 17-atom simulation cells. Also the reference energies of the 65-atom simulation cells used in the Results-section, both with and without TC, are presented in Appendix.  

\section{Results}

\subsection{Bulk silicon}

\begin{figure*}[t]
    \centering
    \includegraphics[scale=.5]{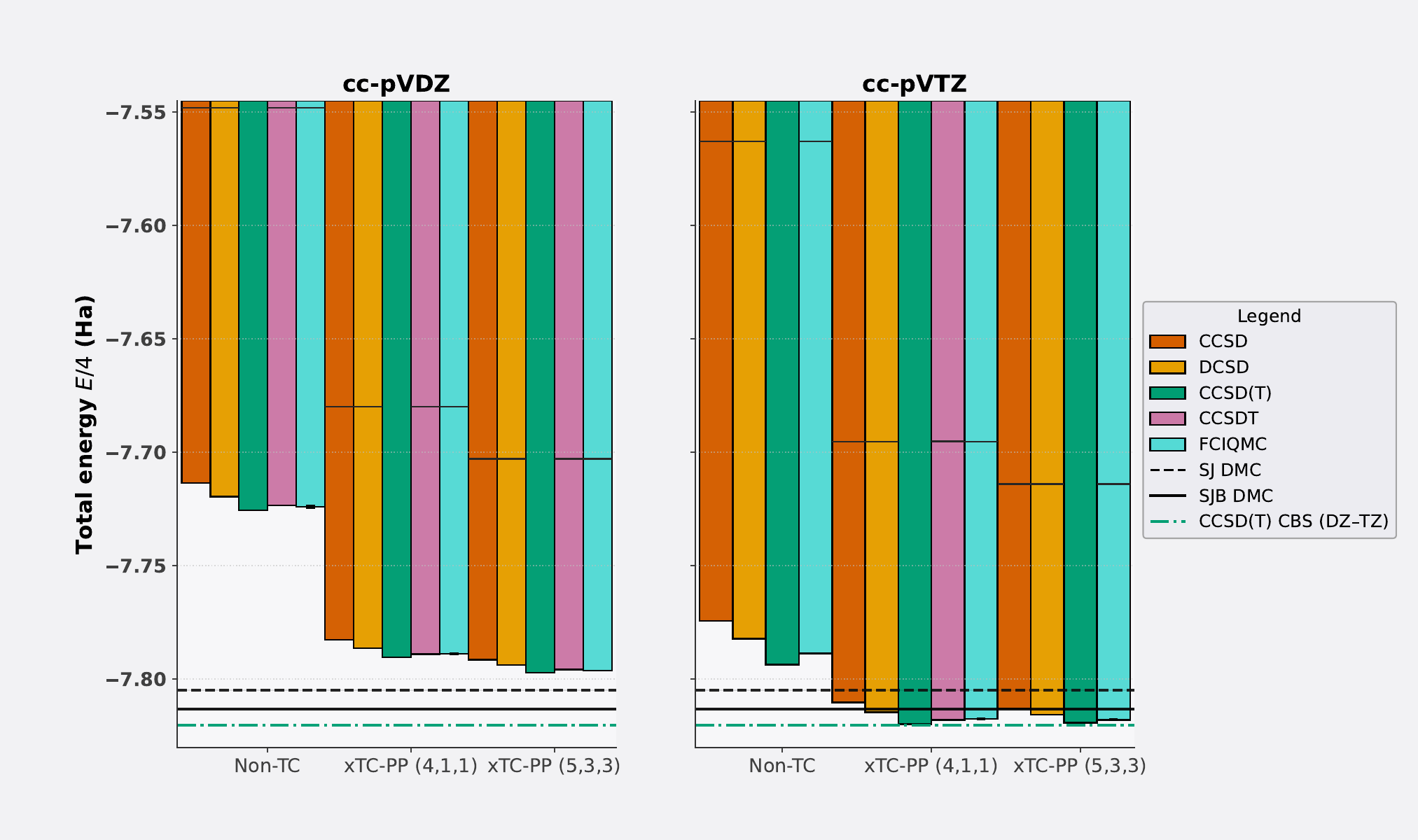}
    \caption{\label{fig: total energies}
    Total electronic energies per primitive unit cell for bulk silicon obtained from eight-atom simulation-cell calculations. 
    Bars show CCSD, DCSD, CCSD(T), and CCSDT results, together with the highest-population FCIQMC (init~10) energies, evaluated using Gaussian DZ and TZ basis sets for the xTC-PP Jastrow cutoff combinations 4\,1\,1 and 5\,3\,3, as well as non-TC cases.
    Horizontal lines indicate fixed-node diffusion Monte Carlo benchmarks with single-determinant (SJ, black dashed) and backflow-corrected (SJB, black solid) trial wave functions; stochastic DMC uncertainty is not visible because it is small. We have also drawn a black solid line accross the bars in the histogram groups to denote the reference energy and the amount of captured correlation energy below the lines. We also show the  extrapolated CSB CCSD(T) estimate for non-TC results as horizontal green dash-dotted line.
    }
\end{figure*}

Figure \ref{fig: total energies} and Table \ref{tab:etot-bulk} report the total electronic energies per primitive unit cell for bulk silicon, obtained from calculations in the conventional eight-atom periodic supercell. We compare coupled-cluster results (CCSD, DCSD, CCSD(T), and CCSDT) with the highest-population FCIQMC energies, and evaluate all of these using Gaussian DZ and TZ basis sets. For the transcorrelated calculations there are two Jastrow-cutoff parameter sets (411 and 533), alongside results obtained with the non-TC Hamiltonian. In addition, we include diffusion Monte Carlo benchmarks, obtained with both Slater-Jastrow (SJ) and backflow-corrected (SJB) forms in the TZ basis. The reference energies are drawn as well. We also show a CBS-extrapolated CCSD(T) estimate.

Figure \ref{fig: total energies} shows that the CC energy decreases through the CCSD-DCSD-CCSD(T) hierarchy. The CCSDT and FCIQMC methods add a small positive correction to CCSD(T) of $\sim1–2$ mH in both DZ and TZ. CCSDT agrees with FCIQMC benchmarks within statistical error except with the 5\,3\,3 DZ Hamiltonian that shows a minor deviation of $\sim0.5$ mE$_h$.

CCSD and DCSD have larger discrepancy with the FCI result than CC with triples, but xTC-PP-CC methods are generally closer to FCIQMC than non-TC CC because of the wave-function compactification. Non-TC CCSD differs from FCIQMC by $4–14$ mE$_h$, xTC-PP-CCSD by $3-8$~mE$_h$. The DCSD is a small over- and CCSD(T) a small underestimation of FCIQMC energy, with deviations up to $3.5$ and $-3$ mE$_h$, respectively, for xTC-PP Hamiltonians. In non-TC case the respective deviations are $6.5$ and $-5$ mE$_h$. As a conclusion we find that xTC-PP-CC with triples, either perturbative or full, provides very accurate results against xTC-PP-FCIQMC.

Examining the reference energies in Fig.~\ref{fig: total energies}, we observe that the primary advantage of the transcorrelated (TC) approach is the substantial improvement in the xTC reference energy over the HF reference energy. The xTC-PP reference energies are systematically lower, reducing the amount of residual correlation energy that the post-HF methods must recover compared to the non-TC Hamiltonian. This improvement becomes more pronounced as the Jastrow factor is enlargened. Nevertheless, the coupled-cluster and FCIQMC methods ultimately bring the total energies into close agreement across different Jastrow choices, as discussed in the next paragraph. 

% Two-column table
\begin{table*}[t]
\centering
\setlength{\tabcolsep}{5pt}
\begin{tabular}{llrrrrrrr}
\toprule
Basis & TC type & CCSD & DCSD & CCSD(T) & CCSDT & FCIQMC & SJ-DMC & SJB-DMC \\
\midrule
DZ & 411     & -7.782690 & -7.786310 & -7.790374 & -7.788940 & -7.7889(3)  &            &             \\
DZ & 533     & -7.791376 & -7.793789 & -7.797095 & -7.795766 & -7.79625(3) &            &             \\
DZ & non-TC  & -7.713525 & -7.719594 & -7.725563 & -7.723366 & -7.7240(5)  &            &             \\
\midrule
TZ & 411     & -7.810257 & -7.814538 & -7.819887 & -7.818081 & -7.8176(2)   & &   \\
TZ & 533     & -7.812840 & -7.815637 & -7.819259 & ---       & -7.8179(2)  & -7.8049(2) & -7.8136(3)  \\
TZ & non-TC  & -7.774302 & -7.782201 & -7.793540 & ---       & -7.78864(3) &  &   \\
\midrule
CBS & non-TC  & ---       & ---       & -7.820187  &  ---       & ---         &  &   \\
\bottomrule
\end{tabular}
\caption{Total energies per primitive 2-atom cell (Ha) based on an 8-atom bulk silicon simulation cell (see also Fig.~\ref{fig: total energies}).}
\label{tab:etot-bulk}
\end{table*}

Figure \ref{fig: total energies} shows that in the DZ basis the two Jastrow factors still yield noticeably different xTC-PP energies ($6$-$7$ mE$_h$). In contrast, the TZ basis removes this sensitivity for high-level methods: xTC-PP CCSD(T) agree to within $<0.6$ mE$_h$ between the two Jastrows. The xTC-PP-FCIQMC energies obtained with the different Jastrows agree exactly. Since any FCI-quality method can differ between similarity-transformed Hamiltonians only through basis-set incompleteness, this agreement demonstrates that TC essentially eliminates basis-set error in TZ.

Because CCSDT is FCI-quality here, and CCSD(T) reproduces it extremely well, their consistency across Jastrow choices provides a strong indicator that the xTC-PP results are at (or extremely near) the CBS limit. This is confirmed by the non-TC CBS CCSD(T) benchmark, which coincides with both TZ xTC-PP-CCSD(T) energies.

These findings make the xTC-PP-CCSD(T) the cheapest reliable FCI-quality method in this context. Larger Jastrow factors may further compactify the wave function, as revealed by the xTC-PP-CCSD and -DCSD results with different jastrow factors, and lower the CC level needed for FCI-quality accuracy. We can see that the xTC-PP-CCSD(T) with 533 Jastrow has a discrepancy with FCIQMC of $1.6$mE$_h$, just within the chemically accurate range, while with 411 the discrepancy that is $0.6$mE$_h$ larger. We leave a systematic study of further wave function compactification with larger Jastrows for future work.

Finally, we compare the xTC-PP results to DMC. The basis set error in DMC can expected to be much smaller than non-TC 2nd quantized DZ and TZ calculations, since the simulation is done in real space using a continuum representation. We can see that the backflow parameterisation yields a notable energy lowering of $\sim 9$mE$_h$ compared to SJ DMC. The SJB energy is $\sim 4.3$mE$_h$ above the xTC-PP-FCIQMC and xTC-PP-CCSDT results in the TZ basis and corresponds roughly to the xTC-PP-CCSD energy in TZ. The fact that backflow introduces a notable correction means that the DMC energy is not converged with respect to wave function complexity. Introducing more variational parameters into the DMC calculation, such as multi-determinant expansions, is expected to lower the DMC energy further. It is difficult to know how much the DMC energy would decrease with a more involved trial wavefunction, but SJB-DMC is often capable of recovering about half of the correlation energy missing at the SJ-DMC level~\cite{rios2006,needs2020}, which would mean SJB-DMC is in agreement with our xTC-PP-FCIQMC energies with an uncertainty of a few mE$_h$.

In conclusion, we find strong indication of basis-set convergence in the transcorrelated calculations with TZ basis sets: CBS extrapolated non-TC CCSD(T) energy  and xTC-PP-CCSD(T) energies with two different Jastrow factors mutually agree. Furthermore, the SJB-DMC energy is also very close to those with the remaining small discrepancy possibly due to the missing correlations caused by the residual fixed node error. We have systematically increased the method accuracy up to FCIQMC and CCSDT levels, finding reliable benchmarks for the correlation treatment. The computationally feasible xTC-PP-CCSD(T) method in TZ basis is found to provide very accurate total energies, with nearly an exact match to FCIQMC and CCSDT benchmarks.

\subsection{Interstitial defects}

In this section we present the formation energies of the silicon self-interstitials, obtained in the $65$-atom supercells with the periodic embedding approach, with both non-TC and xTC-PP coupled cluster approaches. For the sake of completeness, we also report analysis on the magnitudes of basis set and core treatment errors, evaluated in HF level with smaller supercells, in the Appendix. The data in the Appendix also contains the reference energies of the larger supercells, on top of which the CC correlation energies from embedding fragments are added to get the final formation energy estimates.  

\begin{figure*}[t]
    \centering
    \includegraphics[scale=.62]{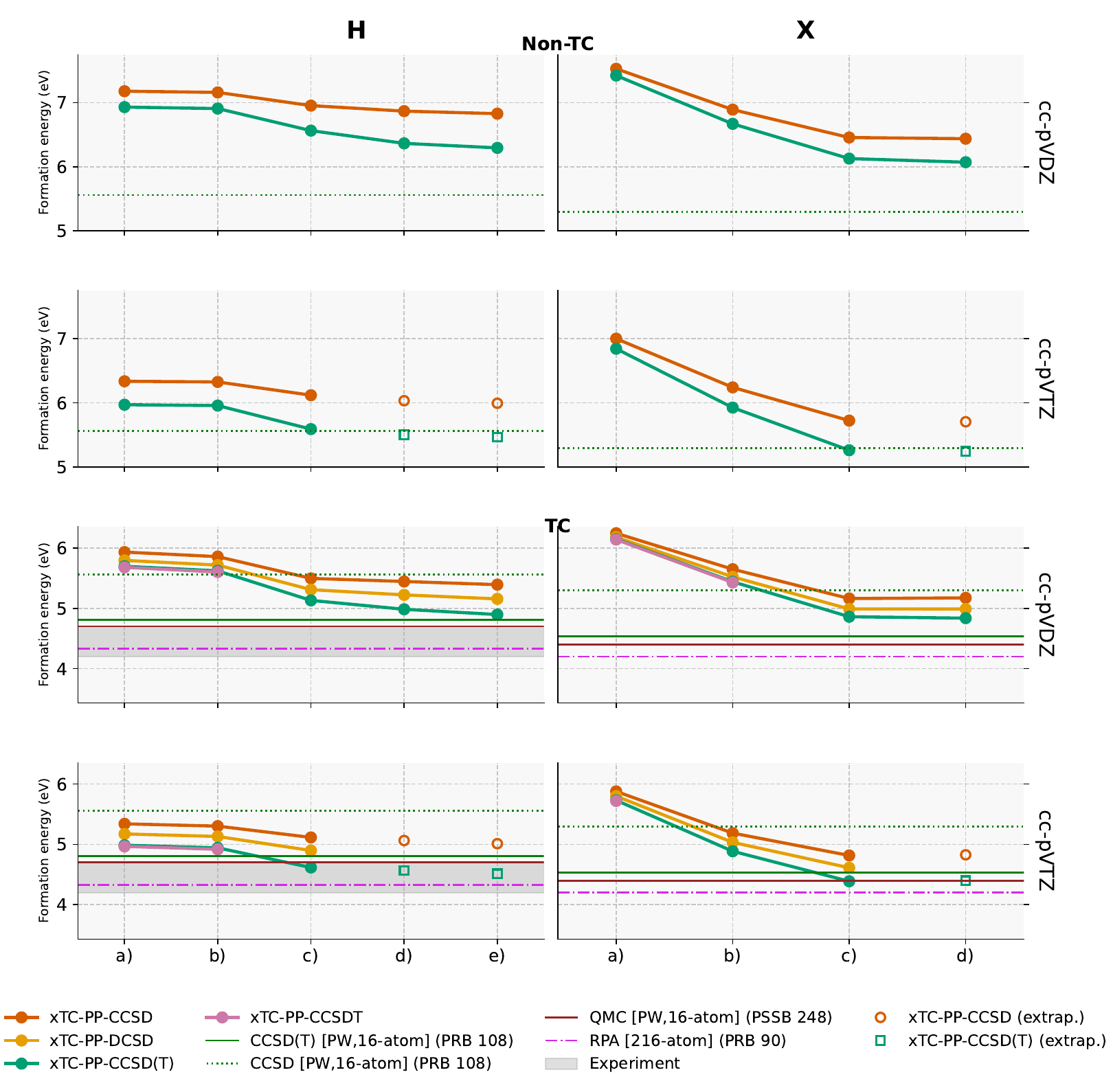}
    \caption{Formation energies of H- and X-interstitials obtained from correlated wave function methods at different basis levels. The upper panels show non-transcorrelated (Non-TC) results, and the lower panels their transcorrelated (xTC–PP) counterparts. Horizontal reference lines indicate periodic benchmark values. Experimental references fall within the grey shaded region for H-interstitial. The x-axis labels (a–e for H, a–d for X) correspond to the fragment pairs defined in Fig.~\ref{figure: fragments}. The Horizontal reference line references are: CCSD\cite{hf-cc-hse}, CCSD(T)\cite{hf-cc-hse}, QMC\cite{qmc} (stochastic errorbar of QMC is omitted for clarity), RPA\cite{rpa-pbe}, HSE\cite{hf-cc-hse}, and PBE\cite{rpa-pbe}.The experimental range of results from different sources\cite{exp1,exp2,exp3,exp4,exp5} is shown by grey shaded region.
    \label{figure: formation_energies}}
\end{figure*}

\begin{figure}
    \centering
    \includegraphics[scale=.4]{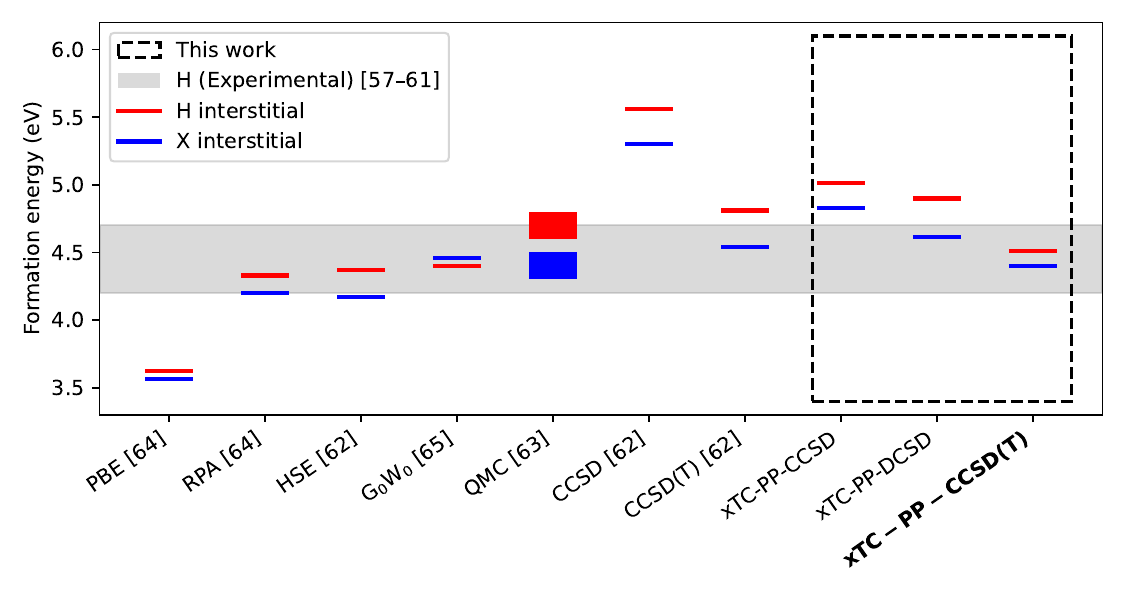}
    \caption{Comparison of largest-fragment formation energies (eV) at the TZ level from this work to previous theoretical benchmarks and experiment. We plot the formation energy from this work and other references as red (H) and blue (X) horizontal lines. Experimental references fall within the grey shaded region for H-interstitial. For the reference values we give the citation numbers in x-labels. References for experimental range are~\cite{exp1,exp2,exp3,exp4,exp5}.}
    \label{figure: formation_energy_comparison}
\end{figure}

Figure~\ref{figure: formation_energies} summarizes the formation energies of H and X interstitials obtained from embedded CC calculations across fragment sizes and basis levels, using both non-TC and xTC-PP Hamiltonians. Also prior benchmark values from theory and experiments are shown, see also Fig.~\ref{figure: formation_energy_comparison} and Table~\ref{tab:formation_pvtz_vs_refs}. In all cases, the formation energies decrease systematically with increasing fragment size, reflecting improved embedding convergence.

Without transcorrelation, the formation energies are substantially overestimated. For the H defect, DZ values of $6.2$–$7.2$ eV lie way above the experimental range of $\sim$4.2–4.7 eV~\cite{exp1,exp2,exp3,exp4,exp5}, while TZ reduces them by about 1 eV but still leaves a noticeable discrepancy. The X defect follows a similar pattern, with DZ values of $6.0$–$6.8$ eV lowered by $\sim$1 eV at the TZ level. Notably, the relative ordering of the two defects reverses with increasing fragment size: X has the higher formation energy for small fragments but becomes lower than H for larger ones, consistent with periodic benchmarks~\cite{hf-cc-hse,qmc,rpa-pbe}.

The non-TC results capture the trends in convergence with respect to fragment size: smaller fragments systematically overestimate formation energies, which approach the periodic supercell limit as fragment size increases. For H, fragments d and e (Fig.~\ref{figure: fragments}) give nearly identical values, and for X, fragments c and d are already converged within a few tens of meV. Because the largest fragments are computationally prohibitive in TZ, we employ an extrapolation scheme that extends each CC flavor using fragment-size corrections from the lower basis set. Because we can do in TZ basis calculations up to fragments (c) for H and X defects, which are found to be very close to convergence in DZ level, the approximation in TZ is expected to be very good. The extrapolated TZ formation energies are reported in Fig.~\ref{figure: formation_energy_comparison} and Table~\ref{tab:formation_pvtz_vs_refs}.

Finally, within the non-TC series, CCSD(T) yields formation energies lower than CCSD by about 0.5 eV for H and 0.3 eV for X, bringing them systematically closer to the benchmark values. This means that inclusion of triple excitations—explicitly or in approximate form—is important.

When we use xTC-PP Hamiltonians for the formation energies, we see a considerable improvement in results, with formation energy reducing from non-TC case by $1.3-1.5$\, eV in DZ and by $\sim 1$\, eV in TZ. This brings the xTC-PP-CC energies into a very good agreement with experimental and theoretical benchmarks. In TZ we again employ the same extrapolation for large-fragment results. In addition to xTC–PP-CCSD and xTC–PP-CCSD(T) results, we also show xTC–PP-DCSD results, which lie between xTC–PP-CCSD and xTC–PP-CCSD(T) values. Also, for smaller fragments we performed full triples xTC–PP-CCSDT calculations, showing that the triples correction on top of to xTC–PP-CCSD(T) energy is negligible. This confirms that xTC-PP-CCSD(T) is capturing all the relevant correlations. 

The formation energy obtained with the reference energies alone is the same for all fragment sizes, as it is evaluated with the full periodic calculation. Comparison of the effect of TC on formation energies in Fig.~\ref{figure: formation_energies} and Table \ref{tab:defect_all} of the Appendix shows that the corrections introduced by xTC-PP are mostly due to correction in the reference energy, before simulating correlations with CC. This is expected as within the TC treatment a large fraction of correlations are included in the reference wave function, and the remaining fraction of correlations to be captured by CC is smaller.

The formation energy convergence with respect to fragment size is found to be similar to the non-TC case. As we are correlating the fragment explicitly with the environment via the xTC-PP Jastrow interaction, this finding is somewhat surprising. Yet, how general this effect is remains to be investigated on more comprehensive benchmarks. Furthermore, it should be noted that with the Jastrow cutoffs of 4\,1\,1 that we use here, the Jastrow correlation length is not very long-ranged (Si bond length is of the order of 4.35 Bohr), which limits the fragment-environment correlation effects to relatively short-range effects. Use of longer-ranged Jastrows can be expected to accelerate convergence to the TDL, and will be pursued in future studies.

The extrapolated largest-fragment TZ results are compared with prior theoretical and experimental benchmarks in Fig.~\ref{figure: formation_energy_comparison} and Table~\ref{tab:formation_pvtz_vs_refs}. Among the methods considered, backflow-QMC~\cite{qmc} (17-atom), RPA~\cite{rpa-pbe} (217-atom), G$_0$W$_0$~\cite{lda-gw} (64-atom), and xTC-PP-CCSD(T) (65-atom) yield H-interstitial formation energies within the experimental range, although due to large stochastic error the agreement of DMC with experiment is unclear. The transcorrelated hierarchy (xTC-PP-CCSD $\rightarrow$ xTC-PP-DCSD $\rightarrow$ xTC-PP-CCSD(T)) shows clear and systematic improvement.

As seen for the total energy in the previous section, a lower-level xTC-PP-CC method, xTC-PP-DCSD, has a relatively good match with SJB DMC. For total energy of the conventional $8$-atom silicon cell, xTC-PP-CCSD had a very good match with SJB-DMC. These results hint that xTC-PP-CC methods with single and double excitations could serve as a cost-effective surrogate to backflow-DMC, while higher levels of transcorrelated coupled cluster or even FCIQMC can be used to improve the accuracy even further.

However, comparison to the theoretical benchmarks should be done with some caution, since most studies have been done in smaller, 17-atom simulation cells as opposed to our embedding calculations in 65-atom supercells. Notably, the G$_0$W$_0$ method was done in $64$-atom supercells\cite{lda-gw}, and yields the best match with our results, although with reversed ordering for the H and X formation energies. 

The effect of the supercell size on the formation energies in $16$, $64$, and $216$ atom supercells has been studied with the random phase approximation (RPA). Their results with $17$- and $65$-atom supercells are equal for the X-defect, and the larger cell brings the H formation energy down by $80$\,meV. This is not enough to match our results with those in~\cite{hf-cc-hse}. The results of the $64$-atom supercells were found to be within $70$\,meV of the results of the $216$-atom supercells for both defects. This indicates that the finite-size errors on formation energies with $65$-atom cells are small, but not insignificant. The use of our xTC-PP embedding method with even larger supercells will be a subject of following studies. 

Based on the studies here and for bulk in the previous section, our results are likely to have very small basis set errors, they are essentially capturing the full wave function correlation effects for these defects, and they are obtained in cells with modest finite-size errors.

\begin{table}[htbp]
\centering
\caption{Largest-fragment formation energies (eV) at the TZ level, and the reference benchmarks.
Values with \dag\ are extrapolated from smaller fragments; letters in parentheses refer to fragment pairs in Fig.~\ref{figure: fragments}. QMC energy is shown with the stochastic errorbar in the last decimal in parenthesis.}
\label{tab:formation_pvtz_vs_refs}
\small
\setlength{\tabcolsep}{5pt}
\renewcommand{\arraystretch}{1.15}
\begin{tabular}{l lccc l}
\toprule
\textbf{Defect} & \textbf{Series} & CCSD & DCSD & CCSD(T) \\
\midrule
\multirow{2}{*}{H} 
 & Non-TC & 5.993\textsuperscript{\dag} (e) & -- & 5.464\textsuperscript{\dag} (e) \\
 & xTC–PP & 5.012\textsuperscript{\dag} (e) & 4.899 (c) & 4.513\textsuperscript{\dag} (e) \\
\addlinespace[3pt]
\multicolumn{5}{l}{
\begin{minipage}{0.92\linewidth}
\footnotesize
\textit{Reference:}
CCSD\cite{hf-cc-hse} = 5.560, CCSD(T) = 4.810\cite{hf-cc-hse}, QMC = 4.7(1)\cite{qmc}, RPA = 4.33\cite{rpa-pbe}, HSE = 4.82\cite{hf-cc-hse}, PBE = 3.626\cite{rpa-pbe}, G$_0$W$_0$ = 4.40\cite{lda-gw};\\
experiment $\sim$4.2–4.7\,eV.
\end{minipage}} \\
\addlinespace[4pt]
\multirow{2}{*}{X} 
 & Non-TC & 5.704\textsuperscript{\dag} (d) & -- & 5.242\textsuperscript{\dag} (d) \\
 & xTC-PP     & 4.827\textsuperscript{\dag} (d) & 4.613 (c) & 4.399\textsuperscript{\dag} (d) \\
\addlinespace[3pt]
\multicolumn{5}{l}{
\begin{minipage}{0.92\linewidth}
\footnotesize
\textit{Reference:}
CCSD\cite{hf-cc-hse} = 5.295, CCSD(T) = 4.535\cite{hf-cc-hse}, QMC = 4.4(1)\cite{qmc}, RPA = 4.20\cite{rpa-pbe}, HSE = 4.46\cite{hf-cc-hse}, PBE = 3.566\cite{rpa-pbe}, G$_0$W$_0$ = 4.46\cite{lda-gw};\\
\end{minipage}} \\
\bottomrule
\end{tabular}
\end{table}

\section{Conclusions and Outlook}

We have developed a transcorrelated wave-function framework for pristine and defected solids. For bulk silicon, fully periodic calculations with xT-PP-CC and xTC-PP-FCIQMC using different Jastrow factors indicate that transcorrelation substantially accelerates basis convergence in Gaussian frameworks. xTC-PP-CCSD(T) in a TZ basis reaches the accuracy comparable to xTC-PP-FCIQMC and seems to improve upon fixed-node DMC benchmarks. For silicon self-interstitials, the transcorrelated embedding formulation yields formation energies that drastically improve upon non-TC calculations, decrease systematically with fragment size and agree well with established periodic references.

Methodologically, there were three most important developmental steps. First, the existing xTC-PP approach needed to be implemented for periodic systems, so that evaluation of Jastrow factors and pseudopotential commutators leveraged minimum-image convention. Second, we found that the numerical integration used to construct the xTC-PP Hamiltonian was greatly accelerated by the use of uniform real-space grids. And finally, the formulation of the transcorrelated embedding scheme was established. The possible implementation of $k$-point sampling in the xTC-PP Hamiltonian remains a topic for future work.

The present study has limitations that point to clear next steps. Our validation focuses on silicon and neutral self-interstitials; broader assessments on ionic, magnetic, and low-symmetry materials—including systems with stronger static correlation—are needed to probe generality. The Gaussian basis sets used here are originally optimized for non-periodic systems~\cite{ccecp2022}, and yet we found near convergent results without linear dependency issues. This sets stage for future studies with basis sets built for periodic systems–possibly optimized directly for use with xTC-PP–for even faster convergence. Also, the Jastrow factors employed here have relatively short correlation lengths; exploring longer-ranged forms may enhance fragment-environment correlations and accelerate convergence to the thermodynamic limit for fully periodic systems. Also, there is some proof in this work that the use of larger Jastrow factors can help to compactify the wave function, allowing lower level CC methods – or even just xTC with the HF reference determinant – to reach good accuracy. The study of the use of much longer-range Jastrow factors is left for future studies.

The periodic mean-field embedding scheme employed in this study has certain limitations for open-shell, charged, or metallic systems, but the TC-embedding framework itself is general and can be combined with alternative embedding strategies. Particularly promising in this respect, at least for non-conducting systems, is the recently introduced aperiodic defect model \cite{aperiodic}. Not only it eliminates the need for expensive large supercell calculations, but also, due to the absence of the unphysical defect replicas, treatment of open-shell and/or charged defects becomes straightforward.

A particular problem in this study was the need of large embedding fragments because of the displaced atoms that should be contained within the fragment. Yet, for simpler defects where the structural relaxation involves only the neighboring atoms, the xTC-PP-embedding may converge already with rather modest fragments. Relatively small fragments may also be sufficient for studying local excitations on defects, as vertical excitation energies are obtained from calculations within a single structure. 

Fully periodic, non-embedded simulations of solid-state systems with xTC–PP Hamiltonians represent another natural direction for further study. While we have shown that xTC-PP–CC with triples included can achieve FCI-quality total energies, the use of xTC-PP Hamiltonians at the MP2, CCSD, or DCSD levels can extend applicability to larger supercells, enabling systematic studies of finite-size effects under transcorrelation.

Overall, our results demonstrate that transcorrelation substantially mitigates basis-set incompleteness in correlated treatments of solids, and that embedding extends these gains to realistic defect cells. By maintaining systematic improvability of correlation treatment while leveraging compact TC Hamiltonians, the approach provides a practical route toward quantitatively reliable wave-function studies of  pristine and defected crystal structures at controlled computational cost.

\acknowledgments
% This part should be discussed with coworkers.
The transcorrelated embedding model and related code was originally formulated and prototyped by Johannes Hauskrecht, and later extended to periodic solid-state defect systems by Kristoffer Simula.

\clearpage

% =======================
% APPENDIX
% =======================
\appendix

\section*{Appendix A: Basis set comparison}
\addcontentsline{toc}{section}{Appendix}
\begin{table*}[t]
    \centering
    \caption{Total and formation energies for different basis sets at the periodic Hartree-Fock level of theory and calculated from the expectation values of the respective xTC-PP-Hamiltonians with the HF wavefunction. AE stands for all-electron, ECP for effective-core-potential, PAW for projector-augmented-wave.}
    \label{tab:defect_all}
    \setlength{\tabcolsep}{4pt}
    \begin{threeparttable}
    \begin{tabular}{ll rrr rrr}
    \toprule
     Family & Basis &
    $E_{\mathrm{bulk}}$ &
    $E_{\mathrm{hex}}$ &
    $E_{\mathrm{X}}$ &
    $E^{\mathrm{form}}_{\mathrm{H}}$ &
    $E^{\mathrm{form}}_{\mathrm{X}}$ &
    $\Delta$ \\
    \midrule

    \textbf{16(17)-atom supercell}, \\ \textbf{$6\times6\times6$ k-mesh, $E_{\rm HF}^{\rm per}$} \\
    \midrule
    Periodic POB & AE/POB-DZVP        & -577.7500 & -4910.5492 & -4910.5603 & 8.87 & 8.56 & 0.30 \\
    Periodic POB & AE/POB-TZVP        & -577.8461 & -4911.3791 & -4911.4096 & 8.53 & 7.70 & 0.83 \\
    Periodic POB & AE/POB-DZVP-REV2\tnote{a}   & -577.7374 & -4910.4486 & --          & 8.69 & --   & --   \\
    Periodic POB & AE/POB-TZVP-REV2   & -577.8777 & -4911.6477 & -4911.6650 & 8.51 & 8.04 & 0.47 \\
    Dunning      & AE/cc-pVDZ\tnote{b}         & -577.9152 & -4911.9625 & -4911.9747 & 8.63 & 8.30 & 0.33 \\
    Dunning      & AE/cc-pVTZ\tnote{c}        & -577.9345 & -4912.1336 & -4912.1433 & 8.44 & 8.18 & 0.26 \\    
    Ahlrichs     & AE/def2-SVP\tnote{d}        & -577.7340 & -4910.4197 & -4910.4341 & 8.70 & 8.31 & 0.39 \\
    Ahlrichs     & AE/def2-TZVP\tnote{c}       & -577.9258 & -4912.0650 & -4912.0747 & 8.30 & 8.03 & 0.27 \\
    Ahlrichs     & AE/def2-TZVPP\tnote{c}      & -577.9259 & -4912.0655 & -4912.0754 & 8.30 & 8.03 & 0.27 \\
    ccECP        & ECP/cc-pVDZ         &  -60.5148 &   -63.9716 &   -63.9859 & 8.86 & 8.46 & 0.40 \\
    ccECP        & ECP/cc-pVTZ         &  -60.6044 &   -64.0867 &   -64.0962 & 8.31 & 8.05 & 0.26 \\
    ccECP        & ECP/PW (E$_{\text{cut}}=1088$eV)             &  -60.6281 &   -64.1137 &   -64.1194 & 8.26 & 8.11 & 0.15 \\
    Ref. \cite{hf-cc-hse}     & PAW/PW (E$_{\text{cut}}=400$eV)              & --         & --          & --          & 8.16 & 7.93 & 0.23 \\
    \midrule
    \textbf{64(65)-atom supercell}, \\ \textbf{$3\times3\times3$ k-mesh, $E_{\rm HF}^{\rm per}$} \\
    \midrule
    ccECP (nonl.) & ECP/cc-pVDZ         & -242.0233 & -245.4765 & -245.4989 & 8.94 & 8.33 & 0.61 \\
    ccECP (nonl.) & ECP/cc-pVTZ         & -242.3970 & -245.8787 & -245.8962 & 8.32 & 7.85 & 0.47 \\
    \midrule 
    \textbf{64(65)-atom supercell}, \\ \textbf{$1\times1\times1$ k-mesh, $E_{\rm HF}^{\rm xTC-PP-per}$} \\
    \midrule
    H Jastrow & ECP/cc-pVDZ         & -246.2987 & -249.8734 & -- & 7.45 & -- & -- \\
    X Jastrow & ECP/cc-pVDZ         & -246.3144 & -- & -249.8994 & -- & 7.17 & (0.27) \\
    H Jastrow & ECP/cc-pVTZ         & -246.6819 & -250.2836 & -- & 6.88 & -- & -- \\
    X Jastrow & ECP/cc-pVTZ         & -246.6965 & -- & -250.3026 & -- & 6.76 & (0.12) \\
    \midrule
    \bottomrule
    
    \end{tabular}
    \begin{tablenotes}
        \item[a]{The SCF for the X-interstitial did not converge}.
        \item[b]{A cutoff threshold for eigenvalues of the overlap matrix of 10$^{-5}$ was used}.
        \item[c]{A cutoff threshold for eigenvalues of the overlap matrix of 10$^{-3}$ was used}.
        \item[d]{A cutoff threshold for eigenvalues of the overlap matrix of 10$^{-4}$ was used}.
    \end{tablenotes}
    \end{threeparttable}
\end{table*}\clearpage

In Table \ref{tab:defect_all}, we have listed the silicon interstitial (and bulk) total and formation energies in 17(16)-atom simulation cells obtained with different gaussian basis sets and with plane-wave basis at the HF level. When looking at the ccECP results, we see a discrepancy of $0.06$ eV between the triple-$\zeta$ and PW results. The discrepancy between the PW calculations, using PAW and ccECP core treatments, is of the order $0.1$~eV. If omitting the POB basis set family, the discrepancies between all the Gaussian-type-orbital valence-triple-$\zeta$ - within both the all-electron and ccECP treatments - and PW results are about 0.2 eV for the formation energies and 0.1 eV for the relative stability. These values give an estimate of the magnitude of errors in HF energy due to basis set incompleteness, approximate core treatment, etc. Although in the previous xTC-PP studies we have found ccECPs to provide very accurate total correlated energies for molecules and atoms~\cite{simula2025ecp,simula2025tc_transition_metals}, the mismatch between different PW calculations, which is of a comparable magnitude, hints that there might be non-negligible errors due to the approximate core electron representation. Nevertheless, for the interstitial formation energies of this study with larger simulation cells, this error estimate is likely an upper bound, since the transcorrelated treatment reduces the basis-set incompleteness error. We also note here that the uncertainty between different experimental estimates of the formation energy of the H-interstitial \cite{exp1,exp2,exp3,exp4,exp5} is about 0.5 eV. Hence with such relatively small discrepancies in the HF results and with the evidence of reduction of the basis set errors with the TC treatment, we expect to be able to provide realistic estimates of the formation energies, at least with the TZ-basis. As concerns the valence-double-$\zeta$ basis set level or the POB-basis sets altogether the deviation between the formation energies is much larger: up to 0.7 eV, which may be too large to be repaired by TC.  

To get the data in Table \ref{tab:defect_all}, we performed a set of test periodic HF calculations for 16-(bulk) and 17-(defect)atom cells with structures taken from Ref.~\cite{hf-cc-hse}. We used both Gaussian and plane-wave (PW) basis set with the ccECPs and calculated the formation energies $E^{form}$ for both defects and their relative stabilities $\Delta$. For these supercells we used the $6\times6\times6$ Monkhorst-Pack $\mathbf{k}$-point grids. The PW calculations were carried out with the Quantum Espresso package~\cite{giannozzi2017}. The tests indicated that the PW cutoff of $1088$eV allows for convergence within $1$\,meV/atom. The results with valence-double-$\zeta$, valence-triple-$\zeta$ and PW basis sets are presented in Table~\ref{tab:defect_all} with those of Ref. \cite{hf-cc-hse} that used PAW for the core treatment and a PW cutoff of $400$eV.
%Table~\ref{tab:defect_all} shows that the formation energies in the $17$-atom simulation cells with Gaussian basis sets differ by $50-60$\,meV from the PW results. 

In this table we also include the reference ccECP energies for larger $64$/$65$-atom cells, calculated using $3\times 3\times 3$ Monkhorst-Pack $\mathbf{k}$-point grids. The structures of these supercells were optimized at the DFT-HSE06/PW level with the fixed experimental lattice parameters. These HF orbitals were also used as a starting point for all the fragment formation energy calculations presented in this paper. 
In Table~\ref{tab:defect_all} we also provide the expectation values for the xTC-PP Hamiltonians with the HF wavefunction for these supercells. The xTC-PP results are obtained with the $\Gamma$-point only. 

The presented results indicate that the deviation in the formation energies due to the larger cell size is rather small. At the same time, the xTC-PP Hamiltonian provides a considerable improvement towards the benchmark values (see Table~\ref{tab:formation_pvtz_vs_refs}) compared to bare HF. 

%We present in Table~\ref{tab:defect_all} also the HF total and formation energies obtained with different all-electron basis sets, and the results from Ref.~\cite{hf-cc-hse}. Overall we find the triple-$\zeta$ basis sets to provide match only approximately to PW results, and the variation between different gaussian basis sets is unsatisfactorily high. 

\section*{Appendix B: xTC-PP integration grids}

\begin{figure*}[!htb]
    \centering
    \includegraphics[scale=.6]{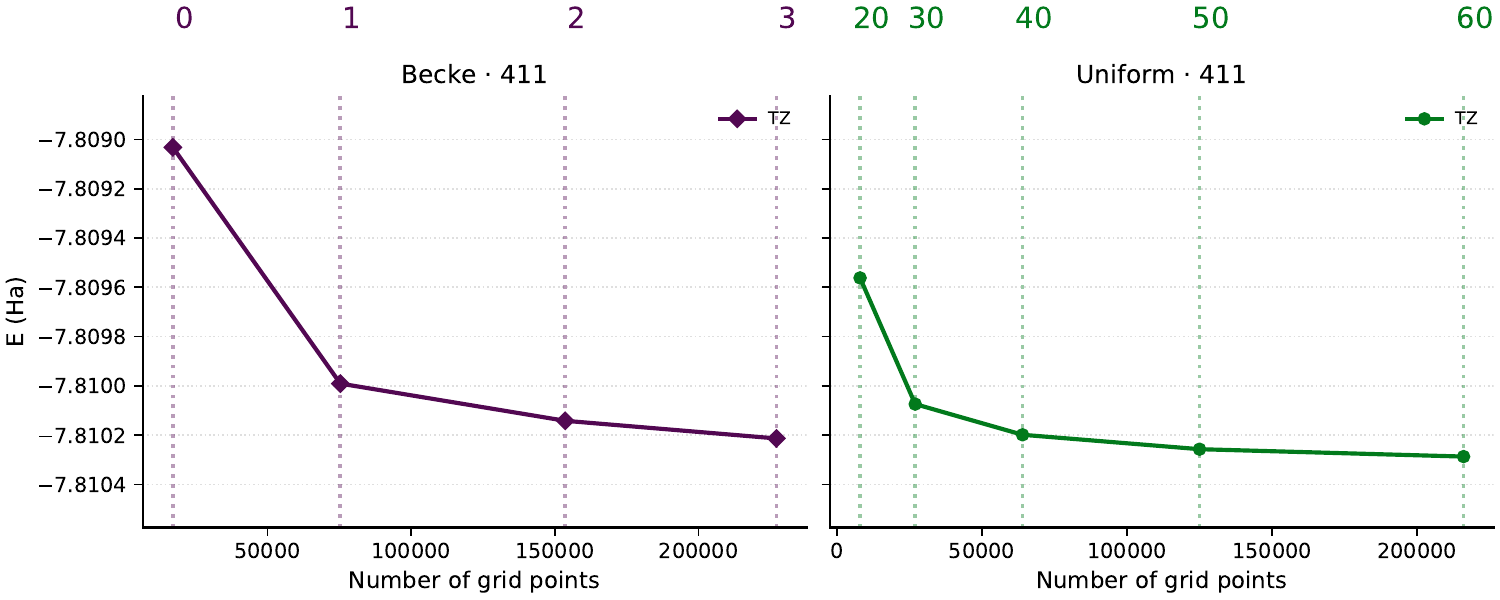}
    \caption{Convergence of the xTC-PP-CCSD energy in an $8$-atom bulk Si cell with respect to the number of integration grid points used for evaluating the xTC-PP integrals. The Jastrow cutoffs are set to $(L_u,L_\chi,L_f)=(4,1,1)$\,bohr. In left, we show the convergence using Becke-type grids obtained from PySCF~\cite{sun2020}, using the grid density levels 0,1,2,3 as implemented in PySCF~\cite{sun2020}. In right, we show the convergence using uniform grids. The Becke grid level is controlled by an integer from $0$ to $3$. The uniform grid is controlled by the number of points along the cartesian axes of the cubic simulation cell. Results were evaluated in TZ basis. }
    \label{figure: tc_integration_grid}
\end{figure*}
\begin{table*}[!htb]
  \caption{TZ, Jastrow 4--1--1: xTC-PP-CCSD reference, correlation, and total energies, evaluated with Si $8$-atom conventional cell, for Becke and uniform grids. $N_{\mathrm{grid}}$ is the total number of integration points; all energies per primitive cell in Hartree.}
  \label{tab:xtc_mp2_grid_tz_411}
  \begin{ruledtabular}
  \begin{tabular}{l r r r r r}
    Scheme & Grid & $N_{\mathrm{grid}}$ & $E_{\mathrm{ref}}$ & $E_{\mathrm{corr}}^{\mathrm{MP2}}$ & $E_{\mathrm{tot}}^{\mathrm{MP2}}$ \\
    \colrule
    Becke   & 0  & 17028  & $-7.692753$ & $-1.222104$ & $-7.809033$ \\
            & 1  & 75320  & $-7.694719$ & $-1.221097$ & $-7.809991$ \\
            & 2  & 153580 & $-7.695026$ & $-1.220940$ & $-7.810142$ \\
            & 3  & 227104 & $-7.695166$ & $-1.220872$ & $-7.810213$ \\
    \colrule
    Uniform & 10 & $10^3 = 1{,}000$    & $-7.685756$ & $-1.226860$ & $-7.806791$ \\
            & 20 & $20^3 = 8{,}000$    & $-7.693926$ & $-1.221460$ & $-7.809562$ \\
            & 30 & $30^3 = 27{,}000$   & $-7.694908$ & $-1.220991$ & $-7.810074$ \\
            & 40 & $40^3 = 64{,}000$   & $-7.695153$ & $-1.220870$ & $-7.810198$ \\
            & 50 & $50^3 = 125{,}000$  & $-7.695254$ & $-1.220827$ & $-7.810257$ \\
            & 60 & $60^3 = 216{,}000$  & $-7.695303$ & $-1.220808$ & $-7.810287$ \\
  \end{tabular}
  \end{ruledtabular}
\end{table*}

\clearpage

%\bibliography{mybib.bib}

\end{document}